\documentclass[conference]{IEEEtran}

\usepackage[utf8]{inputenc}

\usepackage{csquotes}

\usepackage[numbers]{natbib}

\usepackage{graphicx}
\graphicspath{{images/}}

\usepackage{amsmath}

\PassOptionsToPackage{hyphens}{url}\usepackage{hyperref}

\usepackage{array}

\usepackage{multirow}

\usepackage{float}

\usepackage{tcolorbox}

\usepackage{enumitem}

\usepackage{subcaption}

\def\BibTeX{{\rm B\kern-.05em{\sc i\kern-.025em b}\kern-.08em
    T\kern-.1667em\lower.7ex\hbox{E}\kern-.125emX}}

\begin{document}

\title{How Does Microservice Granularity Impact Energy Consumption and Performance? A Controlled Experiment}

\author{
    \IEEEauthorblockN{Yiming Zhao}
    \IEEEauthorblockA{
        Vrije Universiteit Amsterdam\\
        Amsterdam, The Netherlands\\
        y.zhao4@student.vu.nl
    }
    \and
    \IEEEauthorblockN{Tiziano De Matteis}
    \IEEEauthorblockA{
        Vrije Universiteit Amsterdam\\
        Amsterdam, The Netherlands\\
        t.de.matteis@vu.nl
    }
    \and
    \IEEEauthorblockN{Justus Bogner}
    \IEEEauthorblockA{
        Vrije Universiteit Amsterdam\\
        Amsterdam, The Netherlands\\
        j.bogner@vu.nl
    }
}


\maketitle


\begin{abstract}
\textit{Context:} Microservice architectures are a widely used software deployment approach, with benefits regarding flexibility and scalability.
However, their impact on energy consumption is poorly understood, and often overlooked in favor of performance and other quality attributes (QAs).
One understudied concept in this area is microservice granularity, i.e., over how many services the system functionality is distributed.

\textit{Objective:} We therefore aim to analyze the relationship between microservice granularity and two critical QAs in microservice-based systems: energy consumption and performance.

\textit{Method:} We conducted a controlled experiment using two open-source microservice-based systems of different scales: the small Pet Clinic system and the large Train Ticket system.
For each system, we created three levels of granularity by merging or splitting services (coarse, medium, and fine) and then exposed them to five levels of request frequency.

\textit{Results:} Our findings revealed that: i) granularity significantly affected both energy consumption and response time, e.g., in the large system, fine granularity consumed on average 461 J more energy (13\%) and added 5.2 ms to response time (14\%) compared to coarse granularity;
ii) higher request loads significantly increased both energy consumption and response times, with moving from 40 to 400 requests / s resulting in 651 J higher energy consumption (23\%) and 41.2 ms longer response times (98\%);
iii) there is a complex relationship between granularity, system scale, energy consumption, and performance that warrants careful consideration in microservice design.
We derive generalizable takeaways from our results.

\textit{Conclusion:} Microservices practitioners should take our findings into account when making granularity-related decisions, especially for large-scale systems.
\end{abstract}


\section{Introduction}
Microservice architecture has established itself as a popular architectural style in software development~\cite{Newman2015}, and major companies such as Netflix and Amazon have transitioned their applications towards microservices~\cite{7930195}. Unlike traditional monolithic architectures, which often restrict system flexibility, microservices allow different components of a system to be developed using various programming languages, thereby maximizing the advantages of each language.  Microservice architecture has also demonstrated regarding agility, autonomy, scalability, and reusability~\cite{Vale2022}.

While microservices can offer numerous advantages~\cite{Bogner2019}, several important questions warrant further investigation.
Due to the unavoidable transition towards Green IT and Green Software~\cite{Verdecchia2021}, one key concern for microservice-based systems, especially those serving requests at Internet scale, is \textit{energy consumption}.
However, this quality concern has so far received little attention in microservices research, with a recent systematic review by \citet{Araujo2024} including only 37 publications.
Especially concrete architectural decisions have not been studied regarding their impact on energy consumption.

Several important decisions that practitioners must take when developing microservice-based systems are related to \textit{microservice granularity}~\cite{vera2021defining}, i.e., how and over how many services they will distribute the system functionalities.
Allocating a set of operations to many small services results in a more fine-grained system, while using fewer and larger services for the same set of operations leads to a more coarse-grained system.
While the impact of granularity on, e.g., maintainability via dependencies is decently well understood~\cite{Cerny2024}, this is less clear regarding energy consumption.
One hypothesis might be that finer granularity would incur a resource overhead, thereby leading to higher energy consumption, especially as the number of requests fluctuates.
However, coarse-grained services that handle a broader range of tasks may also contribute to increased energy usage due to their complexity.

The relationship between granularity and energy consumption in microservice architectures remains unclear and requires further investigation.
This ambiguity extends to understanding how different levels of request frequency would affect this relationship, as performance and energy consumption are sometimes treated as a trade-off.
These considerations highlight the need for a balanced approach when designing microservice-based systems, considering both performance benefits and potential resource usage.
However, there is a notable lack of empirical evidence on how exactly granularity impacts these quality attributes (QAs).
This knowledge is crucial for microservices practitioners to make informed architectural decisions and to optimize their microservice architectures.

We therefore explore the relationship between microservice granularity and two key QAs: energy consumption and performance efficiency.
As the tech industry increasingly focuses on sustainable computing practices, understanding the energy implications of architectural decisions becomes paramount.
Our research aims to provide empirical evidence to guide software developers in making decisions about microservice granularity when faced with different request workloads that might moderate these relationships.

To address these questions, we conducted a controlled experiment using two open-source microservice-based systems of different scales.
For each system, we created three versions of different granularity (\texttt{coarse}, \texttt{medium}, \texttt{fine}) and subjected them to different request frequencies while measuring energy consumption and response time.
We offer quantitative insights into the trade-offs involved in microservice granularity decisions, and bridge the gap between theoretical discussions of microservice design and practical energy and performance considerations.
Even though our experiment only covered two microservice-based systems, we believe that our findings are decently generalizable and can inform the design of more energy-efficient microservice architectures, aligning with the growing need for sustainable software engineering practices.

\section{Related Work}
Previous research explored various aspects of microservice granularity and its impact on system performance.
\citet{mustafa2017optimizing} investigated the impact of microservice granularity on performance and resource utilization in cloud applications.
Their experiment showed that separating heavily loaded components into distinct microservices improved response times and reduced CPU usage under high load.

Similarly, \citet{10.1145/3147234.3148093} investigated the relationship between microservice granularity and performance.
They compared microservices deployed in a single container versus those partitioned across separate containers.
Their results suggest that services co-located in the same container can benefit from finer granularity, while distributed ones should be designed with coarse granularity for resilience.

Using Service Weaver, which allows developing applications as modular monoliths while deploying them as distributed systems of varying granularity, \citet{galster_performance_2024} evaluated different decompositions of an open-source system.
They concluded that granularity levels beneficial for maintainability had a negative impact on performance.

\citet{7930193} proposed \enquote{microservice ambients}, an architecture meta-modeling approach to address optimal microservice granularity.
Their work extends aspect-oriented ambient modeling with a granularity adaptation aspect.
They demonstrated how the approach supports architecture evolution and runtime analysis.
In a follow-up paper~\cite{hassan_dynamic_2021}, the authors also extended their approach into a runtime adaptation framework to dynamically evaluate different granularity decisions.
While they consider the generic \textit{architectural value} of the different granularity levels, energy consumption is not studied.

Beyond granularity considerations, some researchers have specifically addressed energy efficiency in microservices deployments.
\citet{de2021revisiting} introduced Elergy, a proactive elasticity model for microservices applications that aims to reduce energy consumption while maintaining performance.
Using ARIMA time series forecasting to predict CPU usage for microservice scaling, their model achieved energy savings of 1.93\% to 27.92\% compared to non-elastic scenarios.

\citet{dinga2023empirical} empirically evaluated the energy and performance overhead of monitoring tools on Docker-based microservices.
Their experiments with the Train Ticket benchmark system revealed strong correlations between CPU usage, CPU load, execution time, and energy consumption.
The study provides valuable insights on the trade-offs between monitoring capabilities and system efficiency in microservices.

\citet{9453517} conducted a comprehensive energy consumption analysis of Docker-based Linux distributions for IoT devices.
Using a microservice-based benchmarking architecture with hardware power sensors, they compared the energy efficiency of different operating systems for containerized workloads.
Their results revealed variations in energy consumption patterns, with some systems like RancherOS showing lower power draw but longer task completion times. 

Recent research has also explored hardware approaches to energy optimization.
\citet{9923866} introduced SIMR (Single Instruction Multiple Request), a novel approach for processing microservices that combines GPU-like SIMT execution with out-of-order CPU capabilities.
By executing similar microservice requests, their Request Processing Unit (RPU) achieves 5.7x better efficiency than traditional CPUs.
Their evaluation can significantly reduce both frontend and memory-system energy consumption.

While these studies provide valuable insights into various related aspects, the relationship between granularity and energy consumption remains understudied.
Our research aims to bridge this gap by investigating how different levels of granularity impact both energy consumption and performance.

\section{Study Design}
We used the controlled experiment~\cite{Wohlin2024} research method to fill this gap.
In this section, we describe the details of our experiment design.
As a first step, we used the Goal-Question-Metric (GQM) framework~\cite{caldiera1994goal} to formulate our research goal: 

\begin{center}
    \textit{Analyze} service granularity and request workloads\\
    \textit{For the purpose of} understanding potential trade-offs\\
    \textit{With respect to} energy consumption and performance\\
    \textit{From the view point of} software developers\\
    \textit{In the context of} open-source microservice-based systems\\
\end{center}

Afterward, we derived the following research questions:

\begin{itemize}
    \item \textbf{RQ1:} How do different levels of granularity impact the energy consumption of microservice-based systems?
    \item \textbf{RQ2:} How do different levels of granularity impact the performance efficiency of microservice-based systems?
    \item \textbf{RQ3:} How do rising request workloads impact these QAs in microservice-based systems of various granularity?
\end{itemize}

RQ1 and RQ2 focus on the influence of granularity on the two chosen QAs within microservice architecture, namely energy consumption and performance.
We first conducted experiments with microservice-based systems of varying granularity while measuring the corresponding metrics under uniform workloads, i.e., the number of requests per second sent to the system.
We also compared how granularity affected these QAs across systems of different scales, by choosing one smaller and one larger system.
For RQ3, we focused on examining how the request frequency moderates the potential impact of different granularity levels on the QAs.
We wanted to understand if the impact of one granularity level changes according to the request workload.

\subsection{Experiment Objects}
To investigate potential differences in the impact of granularity, we considered two systems with different sizes (small and large).
For the selection, we used a curated list of open-source microservice-based systems frequently used for research.\footnote{\url{https://github.com/davidetaibi/Microservices\_Project\_List}}
The following inclusion and exclusion criteria guided the selection of our experiment objects.

\noindent \textbf{Inclusion Criteria:}
\begin{itemize}
    \item The system is built with reasonably popular programming languages that are representative of microservices in industry\footnote{\url{https://innovationgraph.github.com/global-metrics/programming-languages}}, e.g., Java, JavaScript, Go, or Python.tem The system can be modified with reasonable effort, e.g., due to modular design and decent code quality.
    This criterion was crucial for adjusting the granularity of the microservices to create the treatments.
    \item The microservice application programming interfaces (APIs) comprehensively cover the entire system functionality and are easily testable.
    A well-defined API facilitates consistent and thorough testing across different granularity levels, ensuring comparability of results.
\end{itemize}
    
\noindent \textbf{Exclusion Criteria:}
\begin{itemize}
    \item The system is a work in progress / incomplete prototype.
    An incomplete system may yield unreliable or non-representative results, potentially skewing our findings.
    \item The system lacks comprehensive documentation, including an incomplete README file and insufficient code comments.
    Documentation is crucial for setup and modifications. This includes both a complete README and commented code for understanding the system structure.
\end{itemize}

Using these criteria as guidance, we selected two suitable systems of different functional scope, namely the Train Ticket system\footnote{\url{https://github.com/FudanSELab/train-ticket/}} and the Pet Clinic system\footnote{\url{https://github.com/spring-petclinic/spring-petclinic-microservices}}. 
In their original versions, Train Ticket contains 40 microservices and Pet Clinic contains 4 microservices.
To create different versions of granularity, we then modified these systems by splitting and merging services.
For three granularity levels, we needed to create two additional versions per system.
This required us to select how to measure system granularity and how to choose the number of services for each granularity level.

No universally accepted assessment methods for measuring microservice granularity exist~\cite{vera2021defining}.
However, two commonly used metrics in the context of granularity are the average source lines of code (SLOC) per service and the average weighted service interface count (WSIC)~\cite{hirzalla2009metrics}.
For our study, we chose WSIC as our primary metric for designing the three different levels of granularity, as it better reflects the complexity of service interactions and is less dependent on implementation details than SLOC.
WSIC assesses the granularity of a service based on the number and weight of the operations exposed in its interfaces.
It is calculated for service $S$ with $n$ interface operations as follows:

\begin{equation}
  WSIC_s = \sum_{i=1}^{n} W_i \tag{1}
  \label{eq:wsic_service}
\end{equation}

where \textit{i} represents each interface operation and $W_i$ denotes its weight.
Weights typically reflect interface complexity, considering factors such as the number and types of parameters.
In our study, we assigned a weight of 1 to all interface operations, which is the default value and effectively leads to WSIC being the number of service operations.

\citet{bogner2020collecting} conducted a mining study to identify real-world value ranges for service-oriented metrics including WSIC based on quartiles.
The WSIC results are summarized in Table~\ref{tab:WSIC}.
Guided by these quartiles, we adopted the following criteria for our three different microservice granularity levels.
Fine-grained systems have 5-8 operations, medium-grained systems have 8-15 operations, and coarse-grained systems have 15-31 operations per service.

\begin{table}[ht!]
    \centering
    \caption{WSIC ranges identified by \citet{bogner2020collecting}}
    \begin{tabular}{lrrrr}
        Quartile & Top 25\% & 25\%-50\% & 50\%-75\% & Worst 25\% \\ \hline\hline
        WSIC &  [5, 8] & (8, 15] & (15, 31] & (31, 1126] \\ 
    \end{tabular}
    \label{tab:WSIC}
\end{table}

We describe the different granularity levels for our two systems in Table~\ref{tab:number_service_system}.
For our adjusted Train Ticket system, the fine-grained (original) version has 40 services, the medium-grained version 20, and the coarse-grained version has 10.
For the adjusted Pet Clinic, we have 4 services in the fine-grained (original) system, 2 for the medium-grained one, and 1 for the coarse-grained one.
When modifying the systems, we took great care to not alter the functional scope.
We only merged existing services, without deleting any operations, and also respected existing inter-service invocations.
For our load testing, we then selected a representative sample of interface operations for the experiment.
Due to Train Ticket's larger scale with 40 microservices, we selected 40 operations to ensure coverage across all services.
For the smaller Pet Clinic system, 10 operations were selected.
The selected operations from both systems include the HTTP methods \texttt{GET}, \texttt{PUT}, and \texttt{POST}.

\begin{table}[H]
    \centering
    \caption{Details of granularity levels per system}
    \begin{tabular}{l | rrr | rrr}
     & \multicolumn{3}{c|}{Pet Clinic} & \multicolumn{3}{c}{Train Ticket}\\
    Granularity & coarse & medium & fine & coarse & medium & fine\\
    \hline\hline
    \# of Services & 1 & 2 & 4 & 10 & 20 & 40\\
     mean WSIC & 20 & 10 & 5 & 25 & 12.5 & 6.25\\
    \end{tabular}
    \label{tab:number_service_system}
\end{table}

\subsection{Experimental Variables and Hypotheses}
Our experiment includes the following variables (see also Table~\ref{tab:variables}).
We consider three \textbf{independent variables}:
service granularity, different systems as experiment objects, and request frequency.
Service granularity reflects how a fixed functional scope, i.e., a given number of service operations, is allocated to services within a system, i.e., a more fine-grained system allocates the same functionality to a larger number of microservices.
We already described above how we decided on the three levels of granularity (\texttt{coarse}, \texttt{medium}, \texttt{fine}).
In this study, the experiment objects are two systems of different sizes.
Here, size refers to the summed up WSIC and SLoC in each system.
Request frequency denotes the number of requests per second sent to the system under study.
We studied five levels of increasing request frequencies.
These are outlined in Table~\ref{tab:factor_and_treatment} and will be explained later in this section.
Our two \textbf{dependent variables} are energy consumption of the system in joules (J) and response time in milliseconds (ms), i.e., the time until a client receives a response from the system.
 
\begin{table}[ht!]
    \centering
    \caption{Independent and dependent experiment variables}
    \begin{tabular}{m{2.5cm} m{1.5cm} m{3cm}}
        Name & Scale & Operationalization\\
        \hline\hline
        \multicolumn{3}{c}{\textit{Independent variables}}\\
        \hline
        Service granularity & ordinal & coarse, medium, fine \\
        Systems & categorical & Train Ticket, Pet Clinic \\
        Request frequency & ordinal & level-1 to level-5\\
        \hline
        \multicolumn{3}{c}{\textit{Dependent variables}}\\
        \hline
        Energy consumption & ratio & Measured with Powerstat\\
        Response Time & ratio & Measured with Locust\\
    \end{tabular}
    \label{tab:variables}
\end{table}

Regarding our hypotheses, we examined the potential influence of granularity on two dependent variables: energy consumption ($e$) and response time ($t$).
In the hypothesis for RQ1 and RQ2, \textit{$g$} represents the granularity of the microservice-based systems and \textit{$d$} corresponds to the dependent variables, with \textit{$d \in \{e,t\}$}.
We tested the two null hypotheses that the energy consumption or response time means of the different granularity levels are the same:

\[H_{0}^{d}: \mu_{coarse}^{d} = \mu_{medium}^{d} = \mu_{fine}^{d}\]
\[H_{a}^{d}: \exists (g_{1},g_{2}) \mu_{g_{1}}^{d} \neq \mu_{g_{2}}^{d}\]
\[\forall g_{1}, g_{2} \in{\{coarse, medium, fine\}} \]

For RQ3, we used multiple linear regression to analyze how the independent variables including request frequency affected energy consumption and response time.
We therefore did not specify a hypothesis pair for RQ3.

\subsection{Experiment Design \& Execution}
Table~\ref{tab:factor_and_treatment} shows the factors and treatments in our experiment.
We used two systems, Train Ticket and Pet Clinic, and explored three levels of granularity and five types of request frequency.
Employing a full factorial design~\cite{Wohlin2024}, we covered all combinations of these factors, leading to 30 experiment configurations (2 × 5 × 3 = 30).
To minimize the potential influence of random confounders, each experiment configuration was repeated 10 times.
Each of these trials lasted 2 min 30 sec.
One full experiment run therefore took 30 × 10 × 2.5 = 750 min or 12 hours 30 min.
To additionally account for temporal variations in network conditions, we conducted the entire experiment three times, each time several days apart.
This approach resulted in a total experimental duration of 2,250 minutes (750 minutes × 3), equivalent to 37.5 hours.
All data from these experiment iterations were used for analysis to average out potential anomalies.

\begin{table}
    \centering
    \caption{Experiment factors and treatments}
    \begin{tabular}{p{2.154cm}p{6cm}}
        Factor &  Treatments\\
        \hline\hline
        Systems & Train Ticket, Pet Clinic\\
        Request Frequency (requests / s) & level-1 (40), level-2 (120), level-3 (200), level-4 (320), level-5 (400)\\
        Granularity & coarse, medium, fine\\
        \hline\hline
    \end{tabular}
    \label{tab:factor_and_treatment}
\end{table}

\begin{figure}[ht!]
    \centering
    \includegraphics[width=\linewidth]{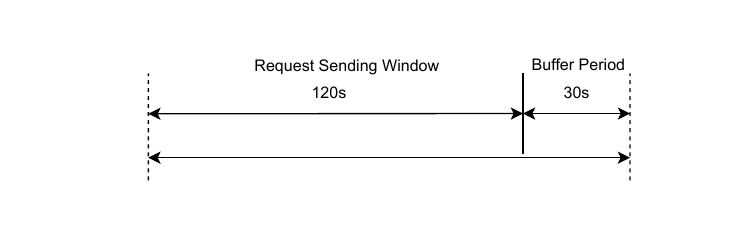}
    \caption{Timeline of one experiment trial}
    \label{fig:timeline}
\end{figure}

Fig.~\ref{fig:timeline} illustrates the timeline of a single trial.
To simulate the differing request workloads, we used the popular load testing tool Locust\footnote{\url{https://locust.io}}, which conveniently offers customizable scripting via Python.
During each trial, Locust continuously issued requests within a designated sending period, depicted as the request sending window in the figure.
Following this active phase, a 30-second buffer period was allocated to allow the server to cooldown.
Otherwise, heat could have a noticeable effect on the energy consumption measurements.

We conducted preliminary experiments with the environment to determine appropriate request frequencies.
We observed that the Train Ticket system exhibited a substantial increase in request failure rates when the frequency exceeded 400 requests per second.
Consequently, we established a range of 0 to 400 requests per second for our experiments.
We defined five load levels (level-1 to level-5) with frequencies of 40, 120, 200, 320, and 400 requests per second, ensuring at least one request per second for each Train Ticket microservice, even at the lowest level.

 tis experiment had two primary componentsExthe perimenthe t Server. 
\begin{itemize}
    \item \textbf{PC}: The personal computer (PC) was equipped with an Apple M1 Pro chip and 16 GB memory, running macOS version 13.4 (22F66). It executed the experiment shell script, which managed the configuration and initiated the Python script that ran the load testing tool, Locust.
    \item \textbf{Experiment Server}: We used an experiment server from VU Amsterdam's Green Lab~\cite{procaccianti_green_2015,mendez_ten_2024}, which is specifically designed for energy efficiency experiments. This server was a Debian GNU/Linux 11 system equipped with an Intel Xeon Silver 4208 CPU @ 2.10GHz and 378 GB of memory. It served two primary functions: measuring energy consumption and providing a platform for system execution via Docker Compose\footnote{\url{https://docs.docker.com/compose}}.
\end{itemize}

\begin{figure}[h]
    \centering
    \includegraphics[width=\linewidth]{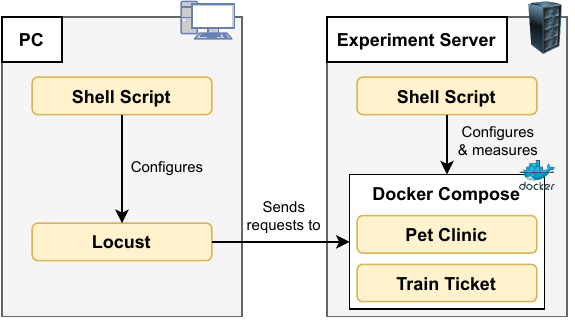}
    \caption{An overview of the experiment infrastructure}
    \label{fig:experiment-infra}
\end{figure}

Fig.~\ref{fig:experiment-infra} provides an overview of the experiment infrastructure.
Before running the experiment, we identified and shut down any non-essential processes running on the experiment server to avoid an impact on our measurements.
Additionally, we monitored the idle experiment environment for a while to ensure that fluctuations in energy consumption, CPU utilization, and memory usage remained within acceptable limits.

We used PowerStat\footnote{\url{https://github.com/ColinIanKing/powerstat}} to measure the energy consumption of the machine using the Intel RAPL (Running Average Power Limit) interface~\cite{Khan2018}.
PowerStat outputs the average power $P$ measured in watts (W) during a period $\Delta t$.
We then calculated the energy consumption $E$ in J using the formula $E = P \times \Delta t$.
We configured PowerStat to collect data at 10-second intervals with a total measurement time of 125 minutes, which is the exact duration of one granularity version for all 5 request frequencies and 10 repetitions.
To prepare the data for analysis, we first identified Locust's start time from its log file, then located the first PowerStat log entry after this time and designated it as the PowerStat start line.
We then used the subsequent 11 lines of PowerStat data for our analysis.
Regarding response time, we simply collected the Locust logs.

\subsection{Data Analysis}
In the first data exploration step, we used descriptive statistics, bar charts, and boxplots to get an overview of the energy consumption and response time.
This gave us a basic understanding of the datasets.
Afterward, we used Q-Q plots and the Shapiro-Wilk test~\cite{razali2011power} to assess the normality of the data.
Q-Q plots should exhibit a linear pattern if the data are normally distributed.
For the Shapiro-Wilk normality test, a p-value greater than 0.05 indicates that the data distribution is not significantly different from a normal distribution.
Examination of the Q-Q plots for both systems across all three granularity levels revealed consistent deviations from the diagonal line, which suggests that the distributions do not conform to normality assumptions.
The Shapiro-Wilk test supported the graphical result, yielding p-values below 0.05 for all datasets, i.e., our dataset was not normally distributed.

Based on this distribution, we chose the Mann-Whitney U test~\cite{mcknight2010mann} for further hypothesis testing.
This non-parametric test evaluates the differences in two datasets using p-values.
A p-value less than 0.05 indicates a significant difference between the datasets, providing evidence for the impact of granularity on QAs.
We used the Holm-Bonferroni correction~\cite{ludbrook1998multiple} to adjust the p-values for multiple comparisons when testing the significance of individual variables.
To analyze the strength of the significant effects, we used Cliff’s Delta ($d$)~\cite{macbeth2011cliff} to see how much each factor affected energy consumption and response time.
Cliff’s $d$ is a non-parametric measure suitable for situations where the data does not conform to a normal distribution~\cite{cliff1993dominance}.
Using a rating scale for Cliff’s $d$, we can easily understand the effect size of each factor, i.e., how strongly it impacts energy consumption and response time.

Since we had multiple independent variables for RQ3, namely granularity, request frequency, and the system, we employed multiple linear regression~\cite{uyanik2013study}.
This allowed us to holistically interpret the coefficients of the independent variables to understand how each contributes to the variation in the dependent variable.
To achieve that, we converted each factor into multiple binary variables based on its levels.
Then, we constructed one linear model to predict energy consumption and one to predict response time.
The model produces p-values to indicate the statistical significance of the contribution of each predictor variable.
Like before, we used the Holm-Bonferroni method to adjust the p-values.

\section{Results}
Our analysis yielded comprehensive insights into the relationship between microservice granularity and the two quality attributes energy consumption and performance.
The complete artifacts, datasets, and analysis scripts are publicly available in the replication package.\footnote{\url{https://doi.org/10.5281/zenodo.14697375}}

For each system under study, we collected 900 distinct measurement groups.
This sample size stems from our factorial design combining 3 granularity levels (\texttt{coarse}, \texttt{medium}, \texttt{fine}), 5 request frequency configurations (40-400 requests / second), 10 repetitions per configuration, and 3 complete experiment iterations.
This design resulted in 900 measurements per system (3 granularity levels × 5 frequencies × 10 iterations × 3 full runs), providing a robust foundation for analysis.

\subsection{Energy Consumption Impact of Granularity (RQ1)}

\begin{figure}[ht!]
    \centering
    \includegraphics[width=0.8\linewidth]{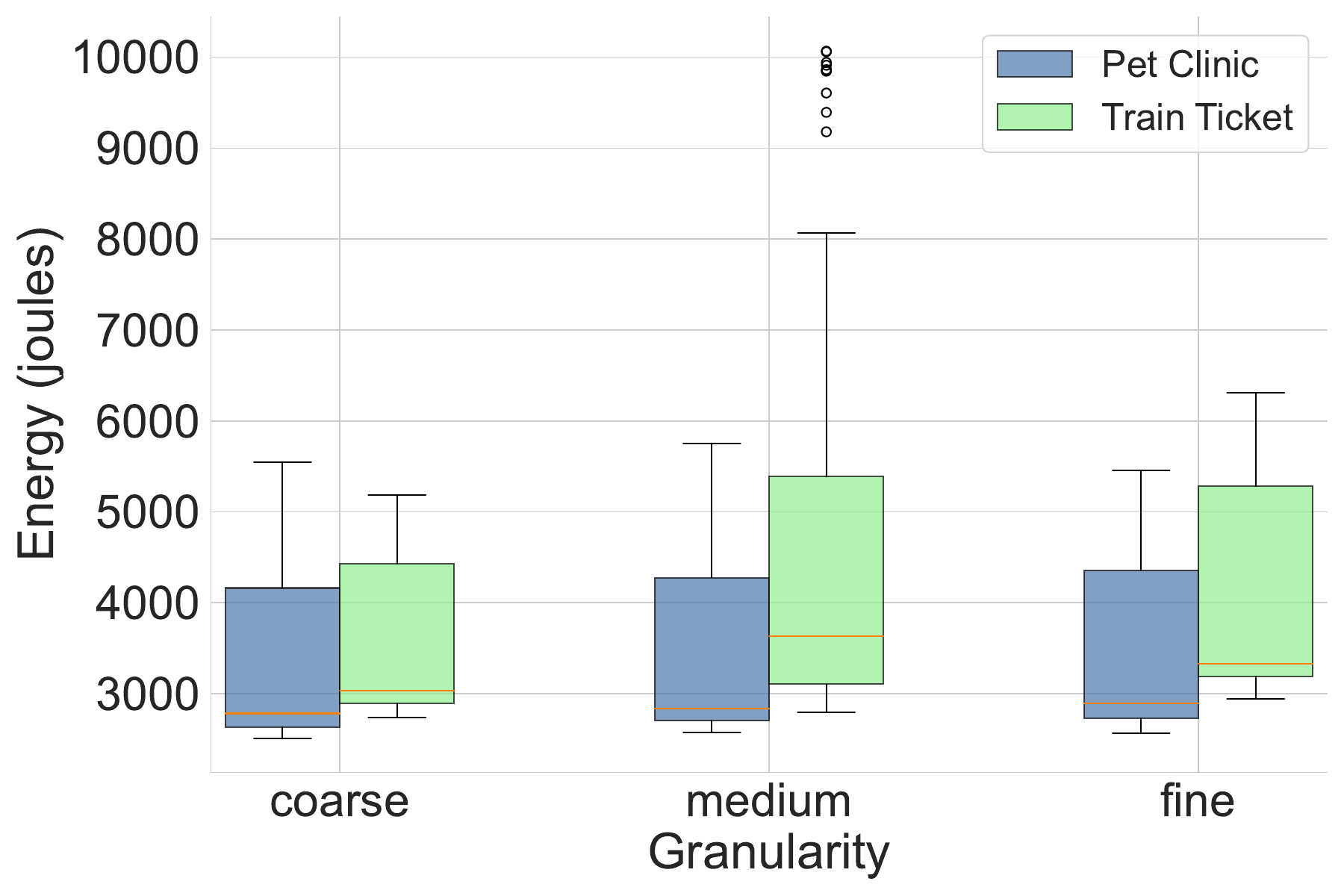}
    \caption{Boxplots for \textbf{energy consumption} (J) of granularities}
    \label{fig:energy_boxplot_rq1}
\end{figure}

Fig.~\ref{fig:energy_boxplot_rq1}  presents the energy consumption distribution for both systems across three granularity levels using boxplots.
A comparative analysis of these plots reveals distinct patterns: Pet Clinic demonstrated relatively consistent behavior with minimal outliers. In contrast, Train Ticket exhibited more variability, with notable outliers for \texttt{medium} granularity.

Table~\ref{tab:descriptive_energy_granularity} provides detailed descriptive statistics for energy consumption, measured in joules, for both systems across all granularity levels.
Pet Clinic showed a clear yet modest increase in mean energy consumption as granularity becomes finer, ranging from 3.25 kJ (\texttt{coarse}) to 3.38 kJ (\texttt{fine}), which is an increase of 4\%.
For Train Ticket, energy consumption was still the lowest at \texttt{coarse} granularity (3.53 kJ), but the peak occurred at \texttt{medium} granularity (4.48 kJ), with \texttt{fine} granularity in the middle (3.99 kJ).
The difference between \texttt{coarse} and \texttt{medium} amounted to nearly 27\% here.
Additionally, the standard deviation indicates higher variability in the Train Ticket system measurements, particularly for \texttt{medium} granularity.
Unsurprisingly, the larger Train Ticket system consistently consumed more energy than Pet Clinic across all levels.

Before continuing with hypothesis testing, we employed the standard Interquartile Range (IQR) method~\cite{dekking2006modern} to remove outliers from both system datasets.
Specifically, we identified outliers as data points falling below $Q1 - 1.5 \times IQR$ or above $Q3 + 1.5 \times IQR$, where Q1 and Q3 represent the first and third quartiles of the energy distributions, respectively, and IQR is the interquartile range (Q3 - Q1).
This approach ensures statistical robustness for our subsequent analysis.

\begin{table}[ht!]
\caption{Descriptive statistics for \textbf{energy consumption} (J)}
\centering
\resizebox{0.5\textwidth}{!}{
\begin{tabular}{l | r r r|r r r}
& \multicolumn{3}{c|}{Pet Clinic} & \multicolumn{3}{c}{Train Ticket} \\
& \multicolumn{1}{c}{coarse} & \multicolumn{1}{c}{medium} & \multicolumn{1}{c|}{fine} & \multicolumn{1}{c}{coarse} & \multicolumn{1}{c}{medium} & \multicolumn{1}{c}{fine} \\ 
\hline\hline
 Mean   & 3250.27 & 3328.89 & 3377.24 & 3527.14 & 4475.92 & 3988.30 \\
 Std    & 808.77  & 836.38  & 863.50  & 852.15  & 1913.66 & 1082.10 \\
 Min    & 2511.84 & 2572.80 & 2564.04 & 2739.84 & 2798.52 & 2946.72 \\
 25\%   & 2634.69 & 2706.24 & 2728.83 & 2895.09 & 3104.82 & 3190.92 \\
 Median & 2783.64 & 2835.36 & 2897.22 & 3031.74 & 3631.86 & 3327.00 \\
 75\%   & 4162.53 & 4276.65 & 4355.97 & 4429.35 & 5388.00 & 5280.84 \\
 Max    & 5544.36 & 5754.72 & 5456.28 & 5188.08 & 10066.08 & 6311.16 \\
\end{tabular}
}
\label{tab:descriptive_energy_granularity}
\end{table}

To understand if these mean differences are statistically significant, we continued with \textbf{hypothesis testing} using the non-parametric Mann-Whitney U test.
Table~\ref{tab:mann-whitney-ene} presents the pair-wise testing results.
When comparing \texttt{fine} and \texttt{medium} granularity, both systems showed p-values exceeding 0.05.
Consequently, we failed to reject the null hypothesis \(H_{0}^{e,g}\), which means that no significant difference in energy consumption was found between these granularity levels.
However, comparisons between \texttt{fine} vs. \texttt{coarse} and \texttt{medium} vs. \texttt{coarse} granularity yielded p-values below 0.05 for both systems. In these cases, we rejected the null hypothesis, which indicates significant differences in energy consumption.

To quantify the magnitude of the significant differences, we employed Cliff's Delta ($d$) for effect size estimation.
The results (see Table~\ref{tab:mann-whitney-ene}) demonstrated varying effect strengths across systems and granularity comparisons.
For the smaller Pet Clinic system, we observed \textit{small} positive effects: 0.221 for \texttt{fine} consuming more energy than \texttt{coarse} and 0.168 for \texttt{medium} consuming more than \texttt{coarse}.
For Train Ticket, we also saw positive effects, but stronger ones: a \textit{large} effect of 0.503 for \texttt{fine} consuming more than \texttt{coarse}, and a \textit{medium} effect of 0.397 for \texttt{medium} consuming more than \texttt{coarse}.
The consistent positive values across these comparisons indicate that finer granularities (both \texttt{fine} and \texttt{medium}) systematically consume more energy than \texttt{coarse} granularity, with this effect particularly evident in the larger Train Ticket system.
However, in both systems, the energy consumption difference between \texttt{fine} and \texttt{medium} was not significant.

\begin{table}[ht!]
\centering
\caption{Pair-wise Mann-Whitney U test results and effect sizes for \textbf{energy consumption} (Holm-Bonferroni-corrected p-values; statistically significant p-values in bold)}
\begin{tabular}{l|rr|rr}
& \multicolumn{2}{c|}{Pet Clinic} & \multicolumn{2}{c}{Train Ticket} \\
& p-value & Cliff’s $d$ & p-value & Cliff’s $d$\\ 
\hline \hline
fine vs. medium   & 0.152 & -- & 0.057 & -- \\ 
fine vs. coarse   & \textbf{0.001} & 0.221 & \textbf{$<$0.001} & 0.503\\
medium vs. coarse & \textbf{0.012} & 0.168 & \textbf{$<$0.001} & 0.397\\ 
\end{tabular}
\label{tab:mann-whitney-ene}
\end{table}

\subsection{Response Time Impact of Granularity (RQ2)}
Initial analysis of response time data revealed significant distributional challenges, with measurements exhibiting substantial variance.
While most observations clustered around 100 ms, extreme cases extended to the 1-2 s range.
These outliers heavily impacted the visual representation of the data, particularly in the boxplot visualizations.
To facilitate an interpretation of the data distribution, we employed the above-mentioned IQR method to identify and remove outliers.

\begin{figure}[ht!]
    \centering
    \includegraphics[width=0.8\linewidth]{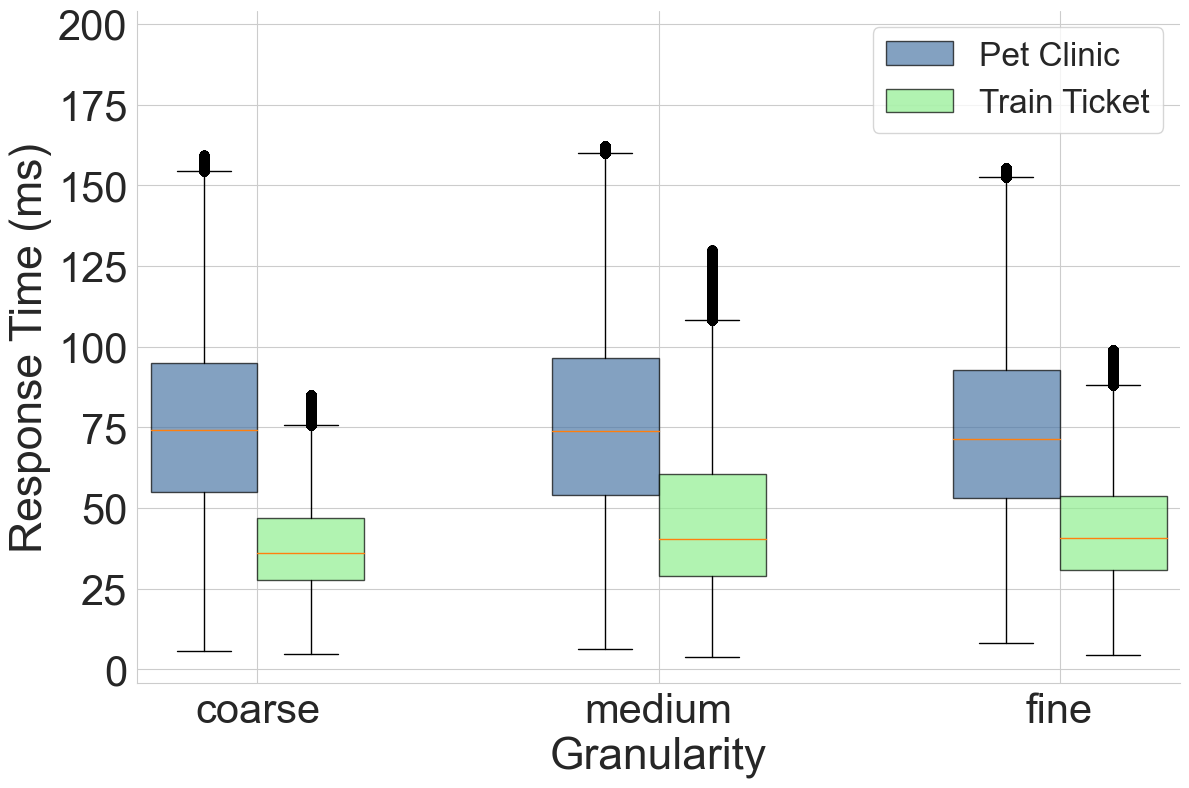}
    \caption{Boxplots for \textbf{response time} (ms) of granularities}
    \label{fig:responsetime_boxplot_rq1}
\end{figure}

Fig.~\ref{fig:responsetime_boxplot_rq1} presents the refined response time distributions after outlier removal.
The Pet Clinic system showed remarkable consistency in response times across all granularity levels, while Train Ticket consistently achieved lower latency compared to Pet Clinic across all granularity configurations.

Table~\ref{tab:response_combined_system_stats} presents the descriptive statistics for the response time of both systems across the three granularity levels.
The analysis of the Pet Clinic system revealed that \texttt{medium} granularity configurations produced the highest average latency (76.44 ms),  followed closely by \texttt{fine} granularity measurements (74.47 ms).
The Train Ticket system demonstrated superior performance overall, with response times ranging from 38.44 ms in \texttt{coarse} granularity to 47.99 ms in \texttt{medium} granularity configurations.
Notably, for both systems, the \texttt{medium} granularity level consistently produced the highest average response times.
Overall, differences here were much less pronounced than for energy consumption, which provides further evidence that the two quality attributes are not always tightly coupled to each other.

\begin{table}[ht!]
    \centering
    \caption{Descriptive statistics for \textbf{response time} (ms)}
        \begin{tabular}{l | r r r | r r r}
    &\multicolumn{3}{c|}{Pet Clinic} & \multicolumn{3}{c}{Train Ticket} \\
   & \multicolumn{1}{c}{coarse} & \multicolumn{1}{c}{medium} & \multicolumn{1}{c|}{fine} & \multicolumn{1}{c}{coarse} & \multicolumn{1}{c}{medium} & \multicolumn{1}{c}{fine} \\ 
    \hline \hline
    Mean   & 76.38 & 76.44 & 74.47 & 38.44 & 47.99 & 43.66 \\
    Std    & 28.28 & 28.79 & 27.66 & 14.96 & 26.26 & 17.54 \\
    Min    & 5.64  & 6.36  & 8.29  & 4.86  & 3.75  & 4.47 \\
    25\%   & 55.03 & 54.06 & 52.95 & 27.54 & 28.91 & 30.81 \\
    Median & 74.08 & 73.75 & 71.43 & 36.00 & 40.35 & 40.59 \\
    75\%   & 94.85 & 96.46 & 92.84 & 46.87 & 60.64 & 53.74 \\
    Max    & 159.28 & 162.21 & 155.52 & 84.91 & 129.83 & 99.07 \\
    \end{tabular}
    \label{tab:response_combined_system_stats}
\end{table}

To understand if these small response time differences were statistically significant, we again employed the Mann-Whitney U tests.
Table~\ref{tab:response_time_mann-whitney-results} presents the results, revealing significant differences across all granularity comparisons ($p < 0.001$).
Based on these findings, we rejected the null hypothesis \(H_{0}^{t,g}\), indicating granularity levels significantly influenced system response times. 
Like last time, we followed up with effect size analysis using Cliff's $d$.
For Pet Clinic, all three significant effects were of very small strength, with the most pronounced difference appearing between \texttt{coarse} and \texttt{fine} granularity (0.040).
In practice, however, all three effects are probably negligible.
Train Ticket, on the other hand, showed slightly more substantial effects, particularly when comparing \texttt{coarse} granularity to \texttt{medium} (-0.158) and to \texttt{fine} (-0.170), which are categorized as small effects.
The difference between \texttt{medium} and \texttt{fine} was again negligible (0.011).
These findings demonstrate that, while granularity significantly affected response times in both systems, the magnitude of this impact ranged from negligible to small.
However, we also saw strong differences between the two systems, indicating that response time considerations of granularity are less important for small-scale systems.

\begin{table}[ht!]
    \centering
    \caption{{Pair-wise Mann-Whitney U test results and effect sizes for \textbf{response time} (Holm-Bonferroni-corrected p-values; statistically significant p-values in bold)}}
    \begin{tabular}{l|rr|rr}
        \multirow{2}{*}{} & \multicolumn{2}{c|}{Pet Clinic} & \multicolumn{2}{c}{Train Ticket} \\
        &p-value & Cliff’s $d$ & -value & Cliff’s $d$  \\ 
        \hline \hline
        coarse vs. medium & $<$\textbf{0.001} & 0.004 & $<$\textbf{0.001} & -0.158 \\ 
        coarse vs. fine   & $<$\textbf{0.001} & 0.040 & $<$\textbf{0.001} & -0.170 \\ 
        medium vs. fine   & $<$\textbf{0.001} & 0.035 & $<$\textbf{0.001} & 0.011 \\ 
    \end{tabular}
    \label{tab:response_time_mann-whitney-results}
\end{table}

\subsection{Energy Consumption Impact of Workload Patterns (RQ3)}
\begin{figure}[ht!]
    \centering
    \begin{subfigure}[b]{0.24\textwidth}
        \centering
        \includegraphics[width=\textwidth]{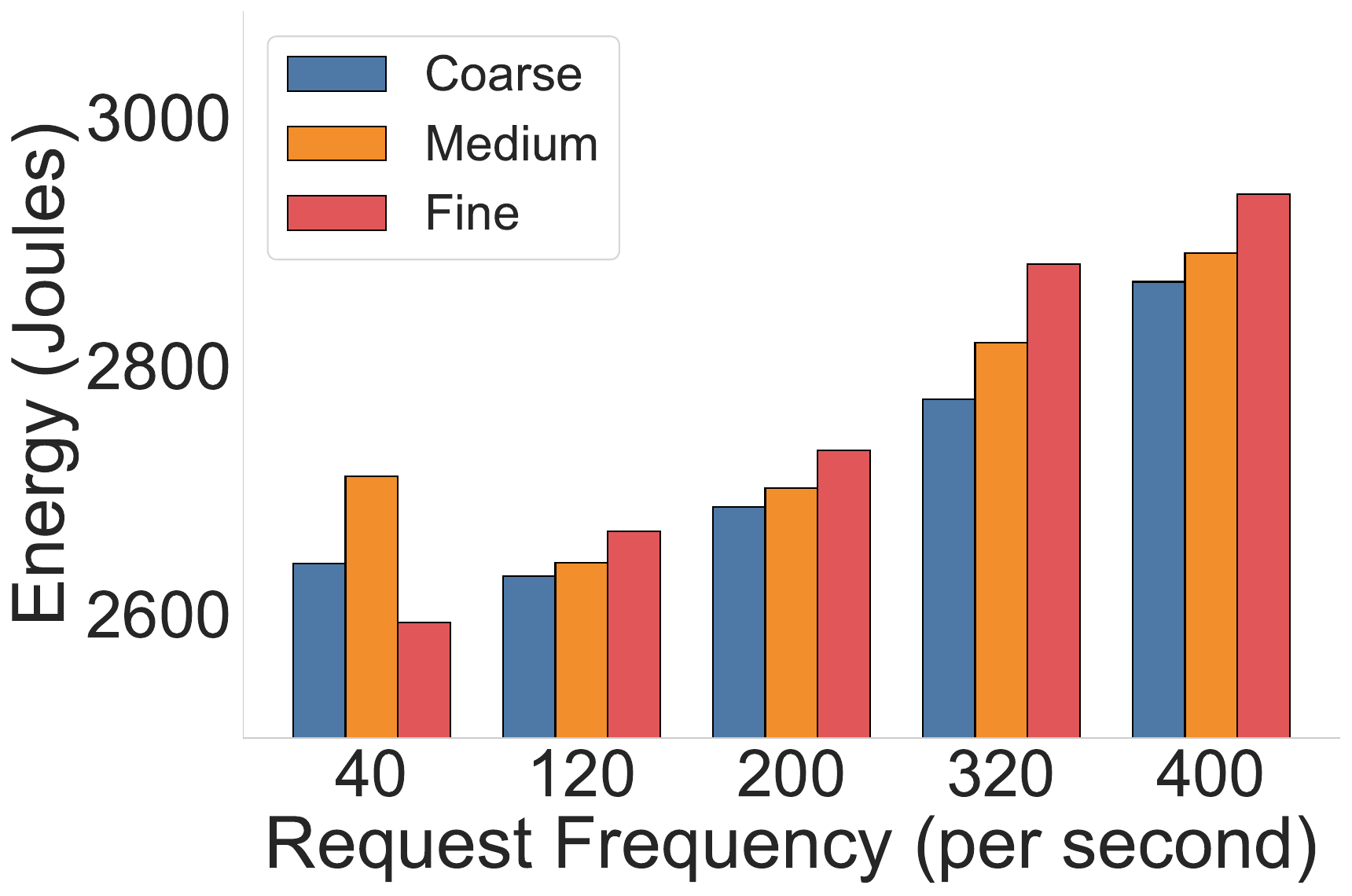}
        \caption{Pet Clinic}
        \label{fig:energy_pet_bar}
    \end{subfigure}
    \hfill
    \begin{subfigure}[b]{0.24\textwidth}
        \centering
        \includegraphics[width=\textwidth]{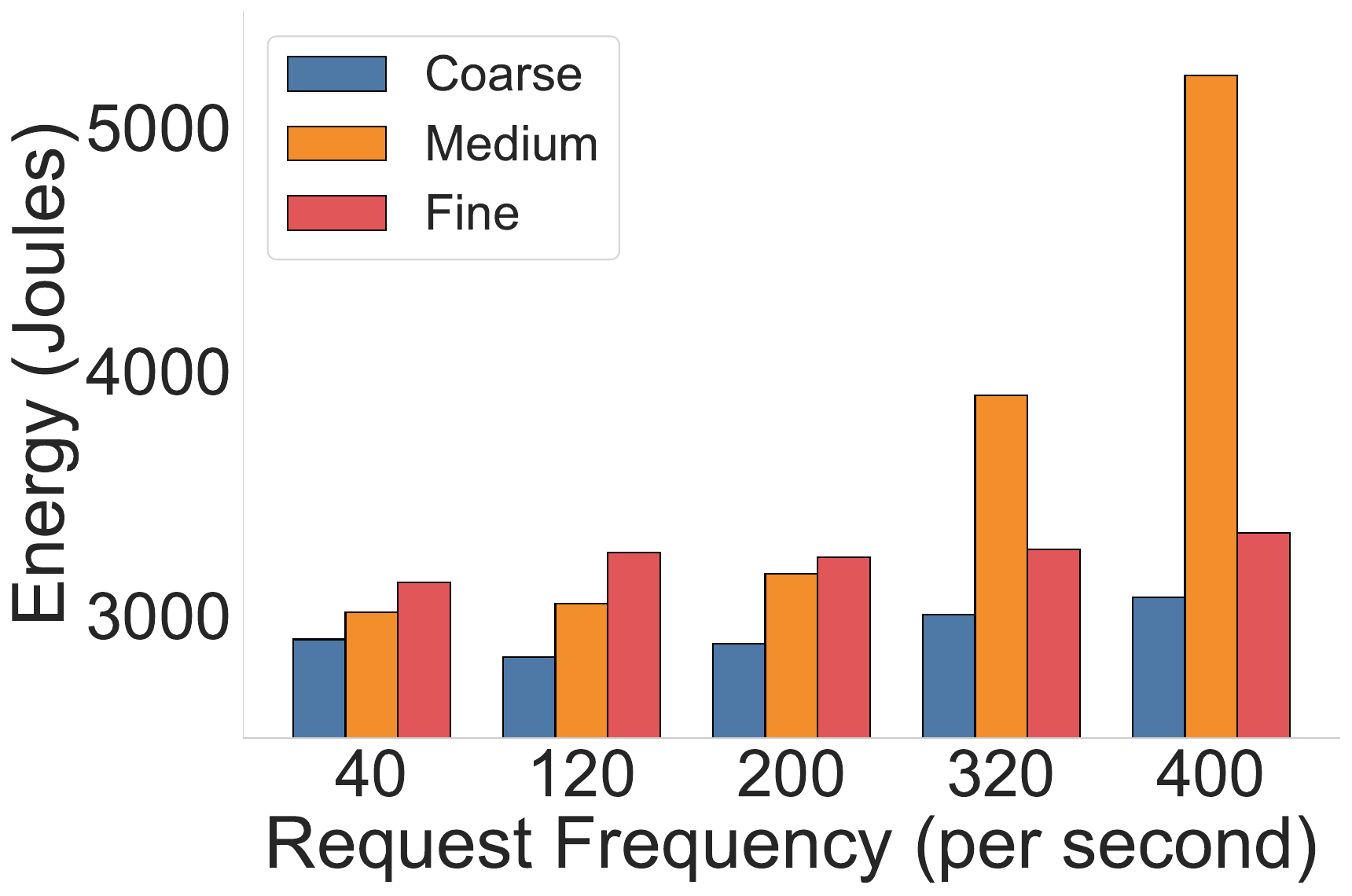}
        \caption{Train Ticket}
        \label{fig:energy_train_bar}
    \end{subfigure}
    \caption{Request frequency impact on \textbf{energy consumption} (J)}
\end{figure}

With RQ3, we wanted to understand how different request frequencies influence the above results.
Regarding energy consumption, Figs.~\ref{fig:energy_pet_bar} and \ref{fig:energy_train_bar} present the average energy consumption by request frequency and system.

For the Pet Clinic system, we saw a consistent trend across most experimental configurations: for each request frequency, energy consumption increases proportionally with finer granularity levels.
The exception to this pattern appears in the 40 requests / s scenario.
Within individual granularity levels, we observe a general upward trend in energy consumption as the request load increases, though this pattern also deviates from the 40 requests / s configuration.

Train Ticket generally followed similar patterns, but it also exhibited notable anomalies, particularly in \texttt{medium} granularity configurations under high load conditions.
Specifically, at 320 and 400 requests / s, we observed unusually high energy consumption values of 3900 and 5300 J respectively, substantially exceeding 3000 J.
Across all configurations, Train Ticket consistently demonstrated higher energy consumption compared to the Pet Clinic system.

To understand the relationship between our various independent variables and the system energy consumption, we performed multiple linear regression.
We used the ordinary least-squares (OLS)~\cite{pedroni2001fully} method for our analysis.
The model includes seven binary parameters: \texttt{is\_gran\_fine}, \texttt{is\_gran\_medium}, \texttt{is\_req\_120}, \texttt{is\_req\_200}, \texttt{is\_req\_320}, \texttt{is\_req\_400} and \texttt{is\_train}, with energy consumption serving as the dependent variable. The regression model explains approximately 14.6\% of the average energy consumption variance (adjusted \(R^2 = 0.146\)), which is decent for such a complex construct.
The F-statistic has a p-value of \(2.31 \times 10^{-23}\), indicating that the model is statistically significant.
We also employed the Holm-Bonferroni method to adjust the p-values to reduce the probability of obtaining Type I errors.

\begin{table}[ht!]
    \centering
    \caption{Multiple linear regression models for energy consumption and response time (Holm-Bonferroni-corrected p-values; statistically significant p-values in bold)}
    \begin{tabular}{l|rr|rr}
    & \multicolumn{2}{c|}{\textbf{Energy Consumption}} & \multicolumn{2}{c}{\textbf{Response Time}} \\
    & p-value & Coefficient & p-value & Coefficient \\
    \hline \hline
    const & \textbf{$<$0.001} & 2816.11 & \textbf{$<$0.001} & 42.44 \\
    is\_gran\_fine & \textbf{0.004} & 293.47 & \textbf{0.048} & 1.80 \\
    is\_gran\_medium & \textbf{$<$0.001} & 513.70 & \textbf{$<$0.001} & 3.41 \\
    is\_req\_120 & 0.85 & 22.50 & \textbf{$<$0.001} & 7.70 \\
    is\_req\_200 & 0.74 & 105.17 & \textbf{$<$0.001} & 16.98 \\
    is\_req\_320 & \textbf{0.004} & 389.58 & \textbf{$<$0.001} & 32.66 \\
    is\_req\_400 & \textbf{$<$0.001} & 650.94 & \textbf{$<$0.001} & 41.18 \\
    is\_train & \textbf{$<$0.001} & 677.93 & \textbf{$<$0.001} & -25.61 \\
    \end{tabular}
    \label{tab:combined_regression_models}
\end{table}

Table~\ref{tab:combined_regression_models} presents the results of the multiple linear regression.
The constant term (2816.11 J) represents the baseline energy consumption for Pet Clinic with \texttt{coarse} granularity and 40 requests / s.
Compared to this baseline, Train Ticket showed a significant increase in energy consumption by 677.93 J, which is the most influential predictor.
However, as expected from the RQ1 results, granularity also played a significant role, with \texttt{medium} granularity (\texttt{is\_gran\_medium}) increasing consumption by 513.70 J and \texttt{fine} granularity (\texttt{is\_gran\_fine}) by 293.47 J.
Request load had a varying impact: high loads of 400 and 320 requests / s significantly increased energy consumption by 650.94 and 389.58 J respectively.
However, lower request loads of 200 and 120 did not show statistically significant effects. 

\begin{figure}[ht!]
    \centering
    \begin{subfigure}[b]{0.24\textwidth}
        \centering
        \includegraphics[width=\textwidth]{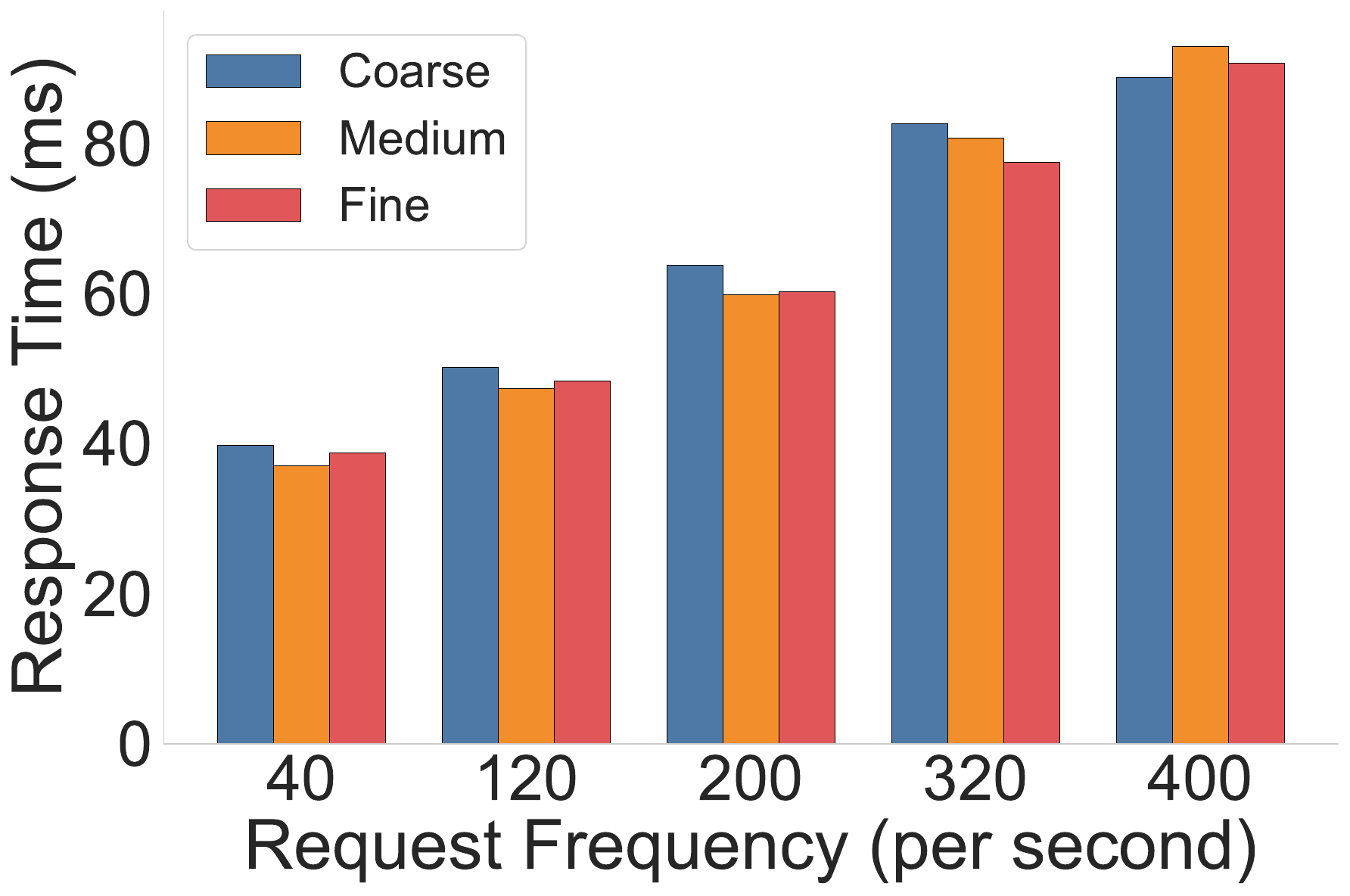}
        \caption{Pet Clinic}
        \label{fig:rt_pet_bar}
    \end{subfigure}
    \hfill
    \begin{subfigure}[b]{0.24\textwidth}
        \centering
        \includegraphics[width=\textwidth]{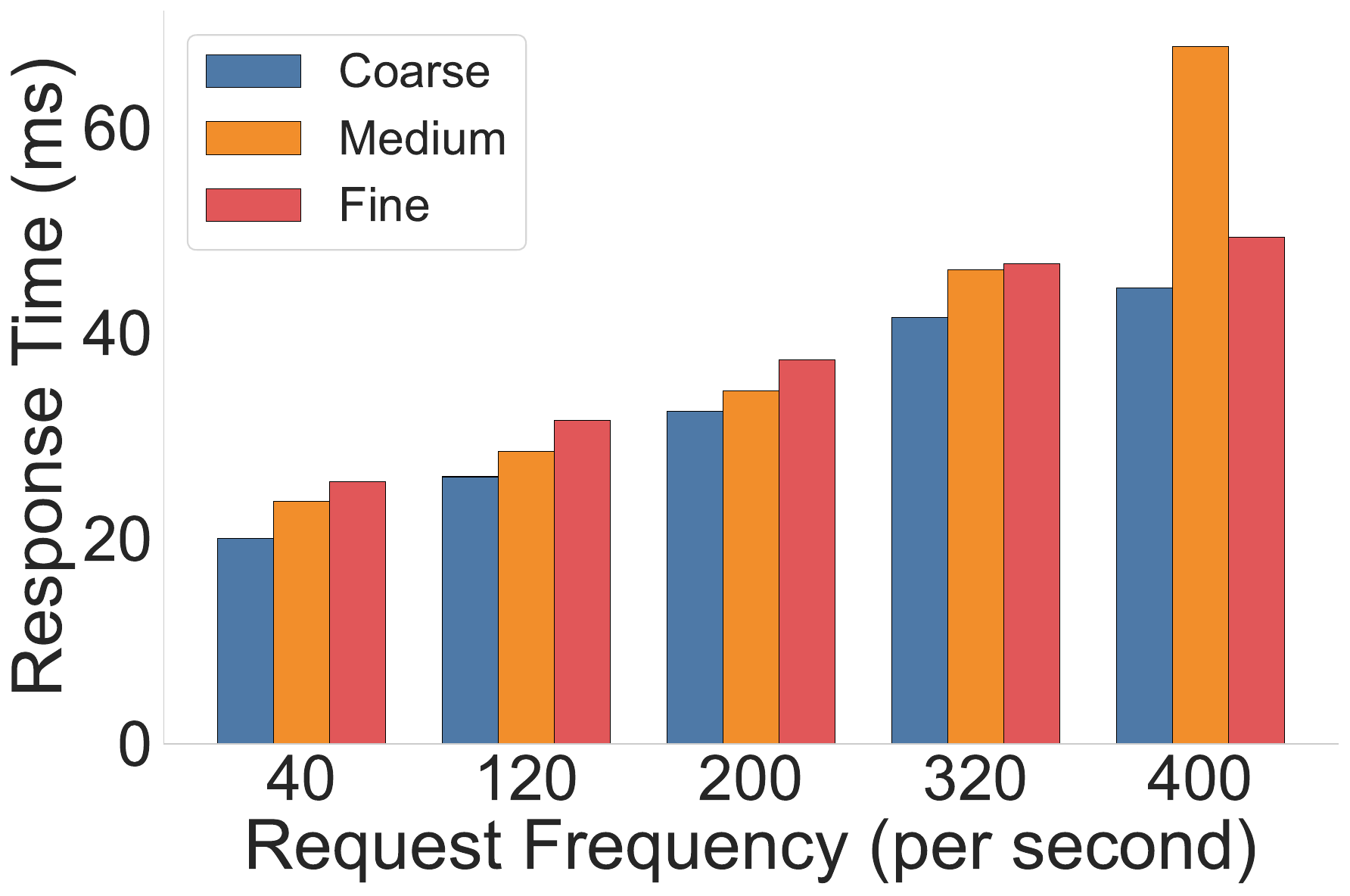}
        \caption{Train Ticket}
        \label{fig:rt_train_bar}
    \end{subfigure}
    \caption{Request frequency impact on \textbf{response time} (ms)}
\end{figure}

\subsection{Response Time Impact of Workload Patterns (RQ3)}
Figs.~\ref{fig:rt_pet_bar} and~\ref{fig:rt_train_bar} present the average response times for all configurations of the Pet Clinic system and the Train Ticket system.
The bar charts for both systems demonstrate a consistent relationship between request frequency and response times:
as request load increased, response times systematically increased across all granularity levels.
The two systems exhibited different response time ranges, with Pet Clinic operating in the 40-100 ms range, while Train Ticket achieved faster responses between 20-80 ms.
At the same request frequency, Train Ticket consistently demonstrated lower response times compared to the Pet Clinic system.

The relationship between granularity and response times varied between the two systems.
In the Pet Clinic system, the relationship between granularity levels and response times showed no consistent pattern.
\texttt{coarse} granularity had the highest response times for the first four request frequencies, but not for 400 requests / s, where \texttt{medium} granularity was the worst.
Additionally, \texttt{medium} granularity had lower response times than \texttt{fine} for the first three levels, but then the relationship switched under higher loads.
In contrast, the Train Ticket system displayed a fairly clear trend: except for \texttt{medium} granularity at 400 requests / s, response times consistently increased as granularity became finer.

We again employed the OLS method to investigate the relationship between our independent variables and response time.
This model incorporated the same parameters as in the energy consumption model.
The only difference is that the dependent variable is response time.
The regression model explains approximately 76.3\% of the variance in average energy consumption (adjusted \(R^2 = 0.763\)), which is very high for such predictions.
The F-statistic has a p-value of \(4.28 \times 10^{-27}\), which is smaller than 0.05, indicating that the model is statistically significant.

Table~\ref{tab:combined_regression_models} presents the results of the multiple linear regression model for system response time.
The constant term (42.44 ms) represents the baseline response time for Pet Clinic with \texttt{coarse} granularity and 40 requests / s.
Compared to this baseline, Train Ticket showed a significant decrease in response time by 25.61 ms, making it the only negative predictor.
Granularity also affected response time, albeit only slightly: \texttt{medium} granularity increased it by 3.41 ms and \texttt{fine} granularity by 1.80 ms.
As expected, request load had a more substantial impact: higher loads progressively increased response times, with 120 requests / s adding only 7.70 ms, but 400 requests / s adding 41.18 ms to the baseline.
These results indicate that, while Train Ticket generally responded faster, increased request loads and both \texttt{medium} and \texttt{fine} granularity lead to longer response times, with request load having the most pronounced effect.

\section{Discussion}
In this section, we summarize our results, provide possible explanations, and discuss the implications for microservices practitioners.
Regarding \textbf{energy consumption}, we generally observed that our test systems consumed more energy as granularity became increasingly finer, i.e., using more services to provide the same functionality requires more energy.
However, there were also some notable exceptions to this general trend.

\begin{tcolorbox}
    \textbf{Main Takeaways for Microservices Practitioners:}
    \begin{enumerate}[leftmargin=0.5cm]
        \item In general, allocating the same functionality over more services will consume more energy.
        \item But: bad decompositions with overloaded services can consume more energy than fine-grained ones.
        \item Energy-related granularity decisions become more important the larger the system and request load.
        \item In general, allocating the same functionality over more services will increase response times.
        \item Without horizontal scaling, energy consumption and response time are not in a trade-off relation.
    \end{enumerate}
\end{tcolorbox}

For example, in the small Pet Clinic system, this relationship did not hold true for the lowest request frequency.
Additionally, the strength of the significant effects was fairly small (just 4\% difference between \texttt{coarse} and \texttt{fine}) and we found no significant difference between \texttt{medium} and \texttt{fine}.
This seems to indicate that energy-related granularity decisions have no major impact for small-scale systems.
Our linear model for energy consumption supports this observation, as the strongest coefficient belonged to the Train Ticket predictor (677.93 J).
There also seems to be a threshold where adding more services does not lead to significantly more energy consumption, most likely because the energy overhead per individual service becomes too small in comparison to the whole system consumption (this was also true for Train Ticket).
In the end, the usage patterns of such systems will determine if energy-related granularity optimizations are worth it.
While 4\% does not seem like much, it can still be substantial energy savings at Internet scale, with millions of concurrent users.
This is also supported by our results that showed the significant impact of high request loads on energy consumption.
In the linear regression model, the higher loads of 320 and 400 requests / s were substantial predictors for increases in energy consumption (389.58 J and 650.94 J respectively).

For Train Ticket, the significant effects were larger, leading to double-digit percentage differences in energy consumption.
This highlights that energy-related granularity decisions play a much more important role in large-scale systems with considerable functionality to distribute.
However, we also saw that the general trend \enquote{more services, more energy} can be easily broken in such systems: at higher request frequencies (320 and 400 requests / s), Train Ticket consumed by far the most energy for \texttt{medium} granularity.
The reason is very likely that the allocation of functionality for \texttt{medium} granularity was not ideal for such high load.
As the respective individual services reached their capacity for serving the incoming requests, their energy consumption and response times skyrocketed.
This leads to an important implication: to make suitable energy-efficient granularity decisions, it is not only important how many services you design, but also how you allocate the functionality to these services.
Overloading individual services can waste more energy than creating additional services to distribute the functionality more evenly.
Interestingly, this overloading was first visible through increased energy consumption (level-4 request frequency) before it noticeably impacted the response time (level-5 request frequency).
This highlights the importance of energy consumption monitoring to identify such overloading.
Response time monitoring alone may detect it too late or not at all if load does not increase beyond the threshold.

Furthermore, horizontal scalability would add another dimension, which was not covered in our experiment. 
For example, \citet{10707516} demonstrated that monolithic systems can be more energy-efficient than microservices when no replicas are used.
However, the situation is reversed with horizontal scaling that reaches optimal configurations regarding availability and response time.
This finding seems consistent with the energy spikes observed for the \texttt{medium} granularity level of Train Ticket.

Regarding \textbf{response time}, we consistently saw significant differences between granularity levels.
In the small Pet Clinic system, \texttt{fine} and \texttt{medium} granularity tended to have lower response times than the coarse-grained monolithic version.
However, these differences were negligibly small, with the largest effect size being $d$ = 0.035.
This indicates that, for small-scale microservice-based systems, granularity-related design decisions have no strong importance for response time and will likely be overshadowed by, e.g., the concrete implementation of the functionality or the used programming language.
However, for the larger Train Ticket system, the opposite was true: with increasingly finer granularity, response times saw significant yet modest increases.
The strongest effect could be observed between \texttt{coarse} and \texttt{fine} granularity, with the latter requiring on average 13.6\% more time per request ($d = -0.170$).
In contrast to the Pet Clinic observations, these results now align with previous studies by \citet{costa2021performance} and \citet{singh2017container}.
The implications of these results are that, for large-scale microservice-based systems, more coarse-grained decompositions generally yield better performance in terms of response time.
As expected, we also saw pretty consistent linear increases in response times with higher request loads, the only exception being the already discussed anomaly of Train Ticket's \texttt{medium} granularity at the highest load level.
These findings highlight the critical role that request frequency plays in system performance, demonstrating that as the system handles more concurrent requests, its ability to respond quickly diminishes proportionally.

Lastly, it is important to note that we did not identify substantial \textbf{trade-offs} between energy consumption and response time, i.e., reducing energy consumption via fewer services did not automatically lead to significantly increased response times.
This was especially not the case in the larger Train Ticket system, where reducing energy consumption and response time via coarser granularity went hand in hand.
For Pet Clinic, saving energy via fewer services indeed meant slightly increased response times, but as mentioned, these differences were negligible outside near real-time systems, for which microservices would be a poor choice anyway.
One important caveat to this is that such a trade-off could indeed manifest with horizontal scaling to accommodate much higher request loads, which was out of scope for this study.

\section{Threats to Validity}
In this section, we discuss several possible threats to validity using the dimensions from \citet{Wohlin2024}.
Regarding \textbf{construct validity}, one potential threat is our used measurement tools.
We used Powerstat for energy consumption measurement, assuming its accuracy and consistency.
Since it is based on RAPL, it includes both CPU and DRAM power.
Preliminary tests with other tools like PowerTop showed consistently lower readings (about 800 J less) than Powerstat.
Both are software-based estimation tools rather than precise measurements and cannot match the accuracy of hardware-based tools like Watts Up Pro used in similar studies~\cite{dinga2023empirical}.
While this may introduce unreliable \textit{absolute} measurements, using software-based estimation tools is an accepted practice in comparative software energy research.
We therefore trust that Powerstat provided consistent \textit{relative} measurements across experimental treatments, allowing valid comparisons.

Another potential threat to construct validity is the definition of microservice granularity.
We used Weighted Service Interface Count (WSIC) as a metric for guiding the creation of different service granularity levels.
However, WSIC may not fully capture all aspects of service granularity.
Alternative metrics such as Source Lines of Code (SLOC) could potentially yield different results.

Regarding \textbf{internal validity}, one potential source for threats is the experiment environment.
We observed that non-consecutive runs produced slightly varying results.
Particularly after server shutdowns, we saw decreases in energy consumption, despite consistent CPU and memory usage. To mitigate this variation, we conducted the experiment three times and averaged the results.
This approach smoothed out fluctuations, providing a more representative measure of the system's energy consumption under different conditions.

Regarding \textbf{external validity}, our study focused on two open-source microservice-based systems of different scales, Train Ticket and Pet Clinic.
While these systems have been frequently used in previous research and differ in complexity, they may not be fully representative, especially for systems from industry.
Different service interdependencies and data flow patterns could lead to slightly different results.
Both systems are also in the realm of online booking and have many operations related to database reads and writes.
Other domains like video or audio processing might have different architectural needs and performance characteristics.

Additionally, the simulated request frequencies in our study may not accurately reflect all real-world usage scenarios.
Due to the design of Locust, requests are sent by a few simulated users at regular intervals.
This does not fully capture the complexity and variability of real-world traffic patterns.

Lastly, regarding \textbf{conclusion validity}, we strengthened our results through a robust experimental design and suitable statistical methods.
We conducted 10 repetitions for each of the 30 trials across factors, repeating the entire experiment 3 times.
This resulted in 900 datasets, providing a substantial sample size that enhances the robustness of our findings and supports strong conclusion validity.
To address the increased risk of Type I errors due to multiple comparisons, we also applied the Holm-Bonferroni method in both the Mann-Whitney U tests and the multiple linear regression analyses.
This approach reduces the likelihood of falsely rejecting null hypotheses, thereby strengthening our conclusion validity.

\section{Conclusion}
This study examined how microservice granularity affects energy consumption and performance in two systems of different scales and under different request frequencies.
Our findings reveal that granularity significantly impacts energy consumption, with more services generally meaning higher energy use.
However, this pattern could be broken through suboptimal service decomposition with overloaded individual services, which would consume more energy than a more fine-grained decomposition.
Regarding response time, granularity had a consistent but small impact, with coarser granularity generally yielding better performance.

These findings have important implications for architecting microservice-based systems.
The choice of granularity should be carefully considered, as it affects both energy consumption and performance.
However, for small-scale systems and low request workloads, the effect of such decisions will likely be overshadowed by more impactful design aspects.
Architects should be aware that reducing energy consumption in microservices does not automatically have to mean that response times need to be sacrificed. 
Our study emphasizes the need for a balanced approach that considers both performance and energy consumption and underscoring the complexity of optimizing microservice architectures.

Future research should explore more diverse systems, more realistic usage patterns, and different technology stacks.
Additionally, analyzing the role of horizontal scaling regarding granularity-related impacts will be important follow-up research.
Studying how service replicas will influence potential trade-offs between energy consumption and response time will be critical in understanding how our results would transfer to high-scalability scenarios.
To easily allow such extensions, we make our artifacts publicly available.\footnote{\url{https://doi.org/10.5281/zenodo.14697375}}

\section*{Acknowledgment}
We kindly thank the Green Lab team at VU Amsterdam for providing and introducing us to the experiment infrastructure used in this study!

\bibliographystyle{IEEEtranN}
\bibliography{references}

\end{document}